\documentclass[12pt]{iopart}
\begin{document}

\letter{Bethe Ansatz
solution of the stochastic process with nonuniform stationary
state.}
\author{A. M. Povolotsky}

\begin{abstract}
The eigenfunctions and eigenvalues of the master-equation for zero
range process on a ring are found  exactly via the Bethe ansatz.
The rates of particle exit from a site providing the Bethe ansatz
applicability are shown to be expressed in terms of single
parameter. Continuous variation  of this parameter leads to the
transition of driving diffusive system from positive to negative
driving through purely diffusive point. The relation of the model
with other Bethe ansatz solvable models is discussed.
\end{abstract}


\address{Bogoliubov Laboratory of Theoretical Physics, Joint Institute for Nuclear Research,
Dubna 141980, Russia}

\ead{povam@thsun1.jinr.ru}

\submitto{\JPA}

%\maketitle

\nosections

 The Bethe Ansatz \cite{bethe} is one of the most powerful
tools to get exact results for the systems with many interacting
degrees of freedom in low dimensions. The exact solutions of
one-dimensional quantum spin chains and two-dimensional vertex
models are classical examples of its application \cite{baxter}.
The last decade the Bethe ansatz has been shown to be useful to
study  one-dimensional stochastic processes \cite{gs,schutz}. The
first and most explored example is the asymmetric simple exclusion
process (ASEP), which serves as a testing ground for many concepts
of the nonequilibrium statistical physics \cite{derrida}. Yet
several other Bethe anstazt solvable models of nonequilibrium
processes have been proposed such as multiparticle-hopping
asymmetric diffusion model \cite{sw}, generalizations of drop-push
model \cite{srb,karim1,karim2}, and the asymmetric avalanche
process (ASAP) \cite{piph}.

All these models has a common property. That is, a system evolves
to the stationary state, where all the particle configurations
occur with the same probability. This property can be easily
understood from the structure of the Bethe eigenfunction. Indeed,
the stationary state is given by the groundstate of evolution
operator, which is the eigenfunction with zero eigenvalue and
momentum. Such  Bethe function does not depend on particle
configuration at all and results in the equiprobable ensemble.
Except for a few successful attempts to apply the Bethe ansatz to
systems with nonuniform stationary state, such as ASEP with
blockage \cite{schutz1} or defect particle \cite{de}, there is
still no much progress in this direction.

In the other hand many interesting physical phenomena like
condensation in nonequilibrium systems \cite{evans}, boundary
induced phase transitions \cite{Krug,sd} or
intermittent-continuous flow transition \cite{piph,pph3} become
apparent from non-trivial form of stationary state, which change
dramatically from one point of phase space to another. Typical
example is the zero-range process (ZRP), served as a prototype of
one-dimensional nonequilibrium system exhibiting the condensation
transition \cite{evans}. While its stationary measure has been
studied in detail \cite{gss}, the full dynamical description is
still absent. The aim of this Letter is to obtain the Bethe ansatz
solution of ZRP. We obtain the eigenfunction and eigenvalues of
the master equation of ZRP with special choice of rates, which is
imposed by integrability requirement. We describe the phase
diagram of the model and show that its particular cases are
equivalent to ASEP, drop-push model and noninteracting diffusing
particles. We also discuss the relation between ZRP and ASAP.

Let us consider the system of $p$ particles on a ring consisting
of $N$ sites. Every site can hold an integer number of particles.
Every moment of time, one particle can leave any occupied site,
hopping to the next site clockwise. The rate of hopping $u(n_{i})$
depends only on the occupation number $n_{i}$ of the site of
departure $i$. The stationary measure of such a process has been
found to be a product measure \cite{evans}, i.e. the probability
of configuration $\{n_{i}\}$ specified by the occupation numbers $%
\{n_{1},n_{2},\ldots ,n_{N}\}$ up to the normalization factor is given by
weight
\begin{equation}
W\left(\{n_{i}\}\right) =\prod_{i=1}^{N}f(n_{i})  \label{weights}
\end{equation}
where $f(n)$ is given by
\begin{equation}
f(n)=\prod_{m=1}^{n}\frac{1}{u(m)}  \label{f(n)}
\end{equation}%
if $n>0$, and $f(0)=1$.

Another way to specify the configuration
of system is to use the set of coordinates of $p$ particles $%
\{x_{i}\}=\{x_{1},\ldots ,x_{p}\}$, rather than occupation numbers
$\{n_1,\dots,n_N\}$, the two ways being completely equivalent. Let
us consider the probability $P_{t}(x_{1},\ldots ,x_{p})$ for $p$
particles to occupy sites $x_{1},\ldots ,x_{p}$ at time $t$. It
obeys the master equation defined by the dynamics described. While
the explicit form of the master equation is too bulky to be
written down explicitly, particular examples will be given below
to illustrate the solution. The main idea of its solution is to
look for eigenfunction in the form of the Bethe ansatz weighted
with the stationary weights of corresponding configurations
(\ref{weights})
\begin{equation}
P_{t}(x_{1},\ldots ,x_{p})=W(\{n_{i}\})P_{t}^{0}(x_{1},\ldots ,x_{p})
\label{solution}
\end{equation}%
where $P_{t}^{0}(x_{1},\ldots ,x_{p})$ is the usual Bethe function defined
as follows
\begin{equation}
P_{t}^{0}(x_{1},\ldots ,x_{p})=e^{\lambda t}\sum_{\{\sigma _{1},\ldots
,\sigma _{p}\}}A_{\{\sigma _{1},\ldots ,\sigma
_{p}\}}\prod_{i=1}^{p}z_{\sigma _{i}}^{-x_{i}}  \label{bethe ansatz}
\end{equation}%
Here the summation is taken over all $p!$ permutations $\{\sigma _{1},\dots
,\sigma _{n}\}$ of the numbers $(1,\ldots ,p)$, and the coordinates of
particles are assumed to be ordered in the increasing order $x_{1}\leq
x_{2}\leq \ldots \leq x_{p}$.

To explain the details we first consider two-particle case subsequently
generalizing it to the case of arbitrary $p$. Without lost of generality we
can define
\begin{equation}  \label{u(1),u(2)}
u(1)\equiv 1,\,\,\,\,u(2)\equiv u>0.
\end{equation}
Then we note that the function
\begin{equation}  \label{p=2,Bethe ansatz}
P_{t}^{0}(x_{1},x_{2})=e^{\lambda t}\left(
A_{1,2}z_{1}^{x_{1}}z_{2}^{x_{2}}+A_{2,1}z_{1}^{x_{2}}z_{2}^{x_{1}}\right)
\end{equation}
satisfies the equation
\begin{equation}
\partial
_{t}P_{t}^{0}(x_{1},x_{2})=P_{t}^{0}(x_{1}-1,x_{2})+P_{t}^{0}(x_{1},x_{2}-1)-2P_{t}^{0}(x_{1},x_{2}),
\label{master02}
\end{equation}%
provided that
\begin{equation}
\lambda =z_{1}+z_{2}-2.
\end{equation}%
When $\left( x_{2}-x_{1}\right) \geq 2$, the same equation
describes the probability $P_{t}(x_{1},x_{2})$ for ZRP. For
$x_{1}=x_{2}-1=x$ the
equation for ZRP%
\begin{equation}
\partial _{t}P_{t}(x,x+1)=P_{t}(x-1,x)+uP_{t}(x,x)-2P_{t}(x,x+1)
\label{P(x,x+1)eq}
\end{equation}%
can be obtained from \ref{master02}) if we define
\begin{equation}
P_{t}(x,x)=\frac{1}{u}P_{t}^{0}(x,x),  \label{P(x,x)}
\end{equation}%
which also follows from (\ref{solution}). It has been noted above that $%
P_{t}(x_{1},x_{2})$ can be expressed trough
$P_{t}^{0}(x_{1},x_{2})$ in the domain $x_{2}\geq x_{1}$.
Therefore, the term $P_{t}^{0}(x,x-1)$, which appears in the
r.h.s. of (\ref{master02}) when $x_{2}=x_{1}=x$, does not carry
any physical content. The freedom in its definition can be used to
make  the last equation for two-particle  ZRP
\begin{equation}
\partial _{t}P_{t}(x,x)=P_{t}(x-1,x)-uP_{t}(x,x)  \label{P(x,x)eq}
\end{equation}%
consistent with (\ref{master02}). If we redefine
$P_{t}^{0}(x,x-1)$ as follows
\begin{equation}
P_{t}^{0}(x,x-1)=(u-1) P_{t}^{0}(x-1,x)-\left( u-2\right)
P_{t}^{0}(x,x), \label{constraint}
\end{equation}
(\ref{master02}) and (\ref{P(x,x)eq}) become equivalent. The
relation (\ref{constraint}) can be thought of as a constraint for
the function $P_t^0(x_1,x_2)$. After substitution of the ansatz
(\ref{p=2,Bethe ansatz})  this constraint results in
 the relation between the amplitudes $A_{1,2}$ and $A_{2,1}$
\begin{equation}
\frac{A_{1,2}}{A_{2,1}}=-\frac{(2-u)-(1-u)z_2-z_1}{(2-u)-(1-u)z_1-z_2}.
\end{equation}
Taking the cyclic boundary conditions,
$P_t^0(x_1,x_2)=P_t^0(x_2,x_1+N)$, into account reduces the
problem to the system of two algebraic equations. The first one is
the following
\begin{equation}
z_1^{-N}=-\frac{(2-u)-(1-u)z_2-z_1}{(2-u)-(1-u)z_1-z_2},
\end{equation}
while the second can be obtained by the change $z_1\leftrightarrow
z_2$.

A generalization of these results to the case of arbitrary $p$
leads us to the following constraint for $P_{t}^{0}(x_1, \ldots,
x_n)$
\begin{eqnarray}
&&\left( u(n)-1\right) P_{t}^{0}\left( \ldots ,\left( i-1\right)
,\left( i\right) ^{n-1},\ldots \right) -  \nonumber \\
&&\sum_{j=2}^{n}P_{t}^{0}\left( \ldots ,\left(
i\right) ^{j-1},\left( i-1\right) ,\left( i\right) ^{n-j},\ldots \right) - \label{p-constraint}\\
&&\left( u\left( n\right) -n\right) \left( \ldots ,\left( i\right)
^{n},\ldots \right) =0  \nonumber
\end{eqnarray}
where $(i)^n$ denotes $n$ successive  arguments equal to $i$ of
the function $P_{t}^{0}(x_1, \ldots, x_n)$. The Bethe ansatz to be
applicable this relation should be reducible to the two particle
constraint (\ref{constraint}). This can be proved by induction. To
this end, we assume that similar relations including $u(k)$ are
reducible for all $k < n$. Then starting from the relation
(\ref{p-constraint}), which includes  rate $u(n)$, we apply
(\ref{constraint}) to every pair $(i,i-1)$ of arguments of the
function $P_t^0(\ldots)$ under the sum and require the result to
be the similar relation including $u(n-1)$. We obtain the
following recurrent formula for  the rates
\begin{equation}\label{u(n)}
  u(n)=1-(1-u)u(n-1)
\end{equation}
which can be solved in terms of $q$-numbers
\begin{equation}\label{u(n)=[n]}
  u(n)=[n],
\end{equation}
where
\begin{equation}\label{q-numbers}
  [n]=\frac{1-q^n}{1-q}
\end{equation}
with $q=u-1$.

Thus, the solution of the master equation for ZRP with special
choice of the rates (\ref{u(n)}-\ref{q-numbers}) is given by the
ansatz (\ref{solution},\ref{bethe ansatz}) with the eigenvalue
\begin{equation}\label{eigenvalue}
  \lambda=\sum_{i=1}^pz_i-p,
\end{equation}
where the numbers $z_i, \,\,(i=1,\ldots,p)$, satisfy the system of
algebraic equations.
\begin{equation}\label{Bethe equations}
  z_i^{-N}=(-1)^{p-1}\prod_{j=1}^p\frac{(2-u)-(1-u)z_j-z_i}{(2-u)-(1-u)z_i-z_j}
\end{equation}

The physics of the model  can be extracted from the analysis of
these equations. This study is beyond the goals of this Letter and
will be published elsewhere. Nevertheless, we can get some
qualitative information about the phase diagram of the model
considering the behaviour of $u(n)$ and particularly  the limiting
cases of parameter $u$ reproducing the models known before. First,
we note that for $u = 2$, i.e. $q = 1$, the $q$-numbers degenerate
into simple numbers, $[n]_{q=1} = n$. Therefore, the rates are
given by $u(n) = n$, which corresponds to the diffusion of
noninteracting particles. The Bethe equtions in this case are
reduced to
\begin{equation}\label{nonint }
  z_j^N=1,
\end{equation}
as is expected in noninteracting case. In the domain $u > 2$  the
rates $u(n)$ grow exponentially with $n$, which   corresponds to
the interaction between particles effectively accelerating free
diffusive motion, i.e. the higher density of particles the faster
is their mean velocity. In the limit $u\to \infty $ the model is
equivalent to $n = 1$ drop push model \cite{srb}, which is
confirmed by the same form of Bethe equations \cite{karim1}. In
the domain $1<u < 2$ the rates $u(n)$ grow monotonously from $u$
for $n=2$ to  $1/(2-u)$ for $n\to\infty$,  resulting in the
interaction between particles slowing down the particle flow. When
$u=1$, all the rates do not depend on the number of particles,
i.e. $u(n) = 1$. This case can be mapped on the ASEP by insertion
of one extra bond before every particle,
\begin{equation}\label{zrptoasep}
\{x_1,\ldots,x_p\}_{\rm ZRP}\to \{x_1+1,\ldots,x_p+p\}_{\rm ASEP}.
\end{equation}
This mapping results in the totally asymmetric ASEP on the lattice
consisting of $N+p$ sites. At last, in the  domain $0<u<1$ the
rates $u(n)$ also saturate to the constant $1/(2-u)$ with growth
of $n$, though oscillating around this value.

We should note, that both ASEP and drop-push model has been
claimed to belong to Kardar-Parisi-Zhang (KPZ) \cite{KPZ}
universality class\cite{gs,sw}. Similarly, we expect that KPZ
behaviour holds through all the driven-diffusive  phases below and
above the point $u = 2$ and this point itself corresponds to
purely diffusive motion. The transition from one  driven-diffusive
(KPZ) phase to another with opposite driving through purely
diffusive point has been studied in context of  KPZ -
Edwards-Wilkinson crossover in partially asymmetric  ASEP
\cite{kim,dm}. It would be interesting to compare the scaling
behaviour with that appearing in our model.

To show the formal relation of ZRP to another model solved by the
Bethe ansatz, we first remind the formulation of the ASAP
\cite{piph}. Consider particles at one dimensional periodic
lattice. If all particles occupy different sites of the lattice,
every particle cam jump to the right neighboring site with a
Poissonian rate 1. When particle comes to already occupied site,
the avalanche starts. It develops by discrete steps according to
the following rules. If there are $n
> 1$ particles at the same site, either all $n$ particles move to the
next right site at the next time step with probability $\mu_n$ or
$n-1$ particles move one step to the next right site with
probability $(1-\mu_n)$, one particle remains unmoved. The
avalanche is assumed to be infinitely fast comparing to the
Poissonian time, i.e. it finishes before any new jumps of
particles from sites other than active site. Now, let us look at
an avalanche from the reference frame related to an active site.
In this reference frame the particles from all sites other than
active site move simultaneously to the left with probability 1. At
every time step one particle from the active site occupied by $n$
particles can also move to the left with probability $1-\mu_n$.
Obviously in this formulation the ASAP is nothing but the special
case of discrete time ZRP. Indeed, the stationary measure of ASAP
has been shown to be a product measure \cite{pph3}. Moreover, the
integrability imposes the same constraint on the toppling
probabilities in ASAP as on  the rates of exit from a site in ZRP
studied in this Letter,  which can be expressed through the
relation $u(n)=1-\mu_n$ for  $u < 1$.
 It is interesting that
the solution of ASAP was based on usual form of the Bethe ansatz,
resulting in the uniform stationary measure. It, however, allowed
the consideration  of the stable configurations of particles
 only, while avalanches played the role of nonlocal interaction
and led to the master equation containing infinite series.

Despite the formal similarity, there is a big difference between
ASAP and ZRP proposed above. Note, that in the ASAP the dynamics
allows at most one active site to exist at any time moment. This
situation corresponds to the inhomogeneous version of ZRP with one
attracting site. Evans \cite{evans} has shown that in this case
the condensation transition always takes place. So it does in the
case of ASAP, corresponding to the transition from the
intermittent to continuous  flow phase. In the other hand, ZRP
studied in present Letter is spatially homogenous, while the rates
$u(n)$ saturate exponentially fast to the constant with growth of
$n$ when $u<2$. Hence, they do not satisfy Evans condensation
criteria, which requires $u(n)$ to decay slower than $2/n$.
However, the large fluctuations for small $u$ are expected,
leading to phenomenon like sharp crossover. It is of interest to
see how such fluctuations affect the relaxation dynamics.  This is
a matter for further investigation of the Bethe ansatz solution.

To summarize, we used the Bethe ansatz weighted with the
stationary weights of particle configurations to solve the
stochastic process with nontrivial stationary state. We found the
eigenvalues and eigenfunctions of the master equation for zero
range process in continuous time. The system of Bethe equations
was obtained. We showed that the Bethe ansatz to be applicable,
the rates of   exit of particle from a site should be expressed in
terms of q-numbers depending on single parameter. For special
values of this parameter the model corresponds to the drop-push
model, ASEP, and noninteracting particles.  The variation of this
parameter was argued to lead to the transition similar to KPZ -
Edwards-Wilkinson crossover in the partially asymmetric exclusion
process. The relation of the model with asymmetric avalanche
process was discussed.

\ack The author is grateful to V.B. Priezzhev for stimulating
discussion. The work was supported by Russian Foundation for Basic
Research under Grant No.03-01-00780.

\end{document}